\documentclass[article]{pasj00}
\SetRunningHead{Mohammed et al.}{Lensing time delays as a substructure constraint in SDSS~J1004+4112} 

\begin{document}
\title{Lensing time delays as a substructure constraint: a
  case study with the cluster SDSS~J1004+4112.} 
\author{Irshad \textsc{Mohammed}\altaffilmark{1,2}, Prasenjit \textsc{Saha}\altaffilmark{1,2}, and Jori \textsc{Liesenborgs}\altaffilmark{3}}
\altaffiltext{1}{Physik-Institut, University of Zurich, Winterthurerstrasse 190,
  8057 Zurich, Switzerland}
\altaffiltext{2}{Institute for Computational Science, University of Zurich,
  Winterthurerstrasse 190, 8057 Zurich, Switzerland}
\altaffiltext{3}{Expertisecentrum voor Digitale Media, Universiteit Hasselt, Wetenschapspark 2, B-3590, Diepenbeek, Belgium}

\email{irshad@physik.uzh.ch}
\maketitle
 
\begin{abstract}
Gravitational lensing time delays are well known to depend on
cosmological parameters, but they also depend on the details of the
mass distribution of the lens.  It is usual to model the mass
distribution and use time-delay observations to infer cosmological
parameters, but it is naturally also possible to take the cosmological
parameters as given and use time delays as constraints on the mass
distribution.  This paper develops a method to isolate what exactly
those constraints are, using a principal-components analysis of
ensembles of free-form mass models. We find that time delays
provide tighter constraints on the distribution of matter in 
the very high dense regions of the lensing
clusters.   We apply it to the cluster lens
SDSS J1004+4112, whose rich lensing data includes two time delays.  We
find, assuming a concordance cosmology, that the time delays constrain
the central region of the cluster to be rounder and less lopsided than
would be allowed by lensed images alone. This detailed information
about the distribution of the matter is very useful for
studying the dense regions of the galaxy clusters which are very
difficult to study with direct measurements.  A further time-delay
measurement, which is expected, will make this system even more
interesting.
\end{abstract}

\section{Introduction}

While $\Lambda$CDM cosmology is a very successful framework, the
underlying nature of both $\Lambda$ (dark energy) and CDM (dark
matter) remains unknown.  The interaction of baryons with both of
these is well understood during the linear-growth era of structures,
but less so when clusters and galaxies start to form.  $N$-body
simulations of CDM give dark-matter distributions which roughly
follows cuspy profiles with a characteristic radius, so-called
well-known NFW profiles \citep{1997ApJ...490..493N}, however
hydrodynamical simulations and other analytic studies show that in
the presence of baryons, NFW is no longer a good fit in the innermost
part of the haloes (see for example \cite{2014arXiv1409.8617S,2014arXiv1410.6826M}).
The distribution of matter in the innermost part of the galaxy
clusters is dominated by the baryonic component, particularly with BCG
and other elliptical galaxies. So, the distribution is different than
that of dark-matter and does not follow NFW profile, which is a very
good fit in the outskirts of the cluster. Due to the high potential
well of galaxies, some dark-matter contracts adiabatically making its
profile steeper at the centre \citep{1986ApJ...301...27B,2004ApJ...616...16G,
2005MNRAS.356..107R}. Generally, the centres of 
the BCG host active nuclei (AGN), which through feedback pushes 
the gas near the centre of the halo to the outskirts 
\citep{2005MNRAS.360..892D,2006Natur.442..539M, 
2013MNRAS.429.3068T, 2013MNRAS.432.1947M}.  For low mass halos 
AGN feedback can push all the gas outside the halo
whereas for high mass halos AGN feedback is not that strong. At the
very transition from big groups and galaxy clusters, the AGN feedback
is strong enough to push some gas outside the halo but not all. These
processes make the centre of the halo very dynamic and redistribute
the matter near the centre of galaxy clusters. It is difficult to
resolve the structures in those high dense regions by direct
observations,  however, strong gravitational lensing (SL) is capable
of resolving those scales \citep{1998MNRAS.294..734A,1991ApJ...383...66H,
2009MNRAS.397..341L,2008A&A...481...65H,2002sgdh.conf...50K,
2014MNRAS.439.2651M,2014ApJ...795...50S}.  The precise
  identification of the multiply-imaged background galaxies/quasars at
  different redshifts can make SL very powerful. However, there are
  still degeneracies that preclude strong constraints on the central
  regions of galaxy clusters.  In this paper we suggest that lensing
  time delays may provide additional information on the central
  substructure.

The idea of measuring time delays in multiply-imaged lensed systems
was discussed theoretically long before any had been discovered.
\cite{1964MNRAS.128..307R} and \cite{1966MNRAS.132..101R} notably brought some
remarkable insights, which we may summarise as follows.  First, the
time delays due to a lens of mass $M$ is of order $GM/c^3$, hence
weeks to years for galaxy and cluster masses, which is conveniently
human-scaled.  Second, whereas the image data on a lens are all
angular quantities and hence dimensionless, a time delay introduces a
dimension, which in fact is proportional to the Hubble time.  Third,
since lensing depends on the ratio of source-lens and observer-lens
distances, and these distances depend on the cosmological model, time
delays coming from different redshifts can potentially measure the
cosmological parameters.

But lensing time delays also depend on the mass distribution of the
lens, and this introduces uncertainty.  To measure cosmological
parameters one needs a strong prior \citep{2014MNRAS.437..600S},
especially if only a single lens is used \citep{2014ApJ...788L..35S}.
If only the Hubble time is sought, while the $\Omega$ parameters are
assumed, an ensemble of lenses gives better-constrained results
\citep{2006ApJ...650L..17S,2007ApJ...660....1O,
  2008ApJ...679...17C,2010ApJ...712.1378P,2014arXiv1404.2920R}, but
still not as precise (so far) as from the CMB.

In this paper, we reverse the traditional process, and take the
cosmological parameters as given.  Now, in lenses with sources at only
one redshift and given cosmological parameters, a time delay breaks
the steepness degeneracy.  But in situations like SDSS J1004+4112,
where the steepness is already partly constrained by other lensing
observables, time delays provide information about the shape of the
mass distribution, particularly very close to the centre, which is
very difficult to probe with direct observations. These central
regions are also very difficult to probe with weak lensing or flexion
data.
We develop a method to quantify what time-delay measurements tell us
about a lensing mass distribution.  We then apply the method to SDSS
J1004+4112, which has two measured time delays.  A further time delay
is expected, so the results and interpretation are preliminary.
Nonetheless, they provide insight into what may be possible.

\section{The cluster SDSS J1004+4112}

\begin{figure}
\centering
\includegraphics[width=0.45\textwidth]{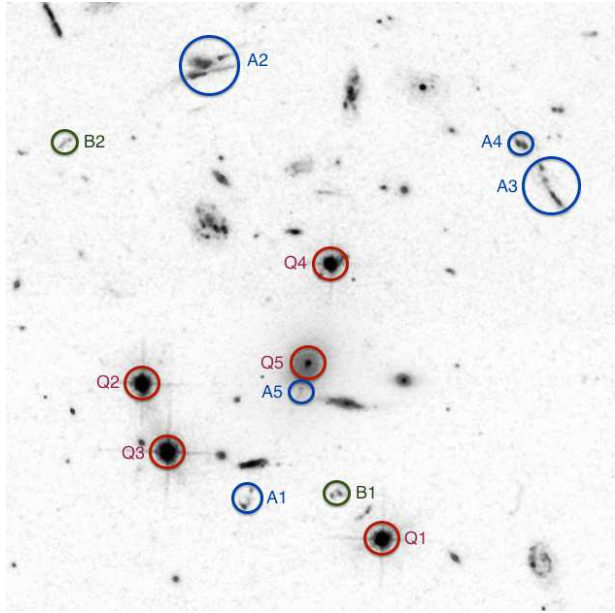}
\caption{Multiple images of the quasar, labelled Q1 to Q5 (red) in order of
  time delay (measured or expected). Showing galaxy A and B images in blue and green respectively. }
\label{fig:raw}
\end{figure}

The SDSS cluster J1004+4112 at redshift 0.68 has three
strongly-lensed systems.  At redshift 1.74, there is a quasar (Q)
lensed into five images (Q1-Q5)
\citep{2003Natur.426..810I,2005PASJ...57L...7I}.  Further, there is a galaxy (A) at redshift 3.332
lensed into five images, and another galaxy (B) at redshift 2.74 is
lensed into two images \citep{2005ApJ...629L..73S} (see Figure
\ref{fig:raw}).  Still another
candidate lensed system (C) is known, but not yet confirmed
spectroscopically.

The Q system is natural for time-delay measurements.  Two of the four
possible time delays (between Q1-Q2 and Q1-Q3) of the quasar images
have been measured \citep{2007ApJ...662...62F,2008ApJ...676..761F} and
a third is expected.  The image separation is large (up to $14''$),
and since time delays scale with the square of the image separation,
the time delays are much longer than with galaxy lenses.  Image Q3
lags the nearby Q2 by 40 days and lags Q1 by 821 days.  The cluster
gas has also been observed in X-rays \citep{2006ApJ...647..215O}.  As
data have accumulated, many different models have been published
\citep{2004ApJ...605...78O,2004AJ....128.2631W,2006PASJ...58..271K,
  2007ApJ...663...29S,2008PASJ...60L..27I,2009MNRAS.397..341L,2010PASJ...62.1017O}.

\begin{figure*}
\centering
\includegraphics[width=0.49\textwidth]{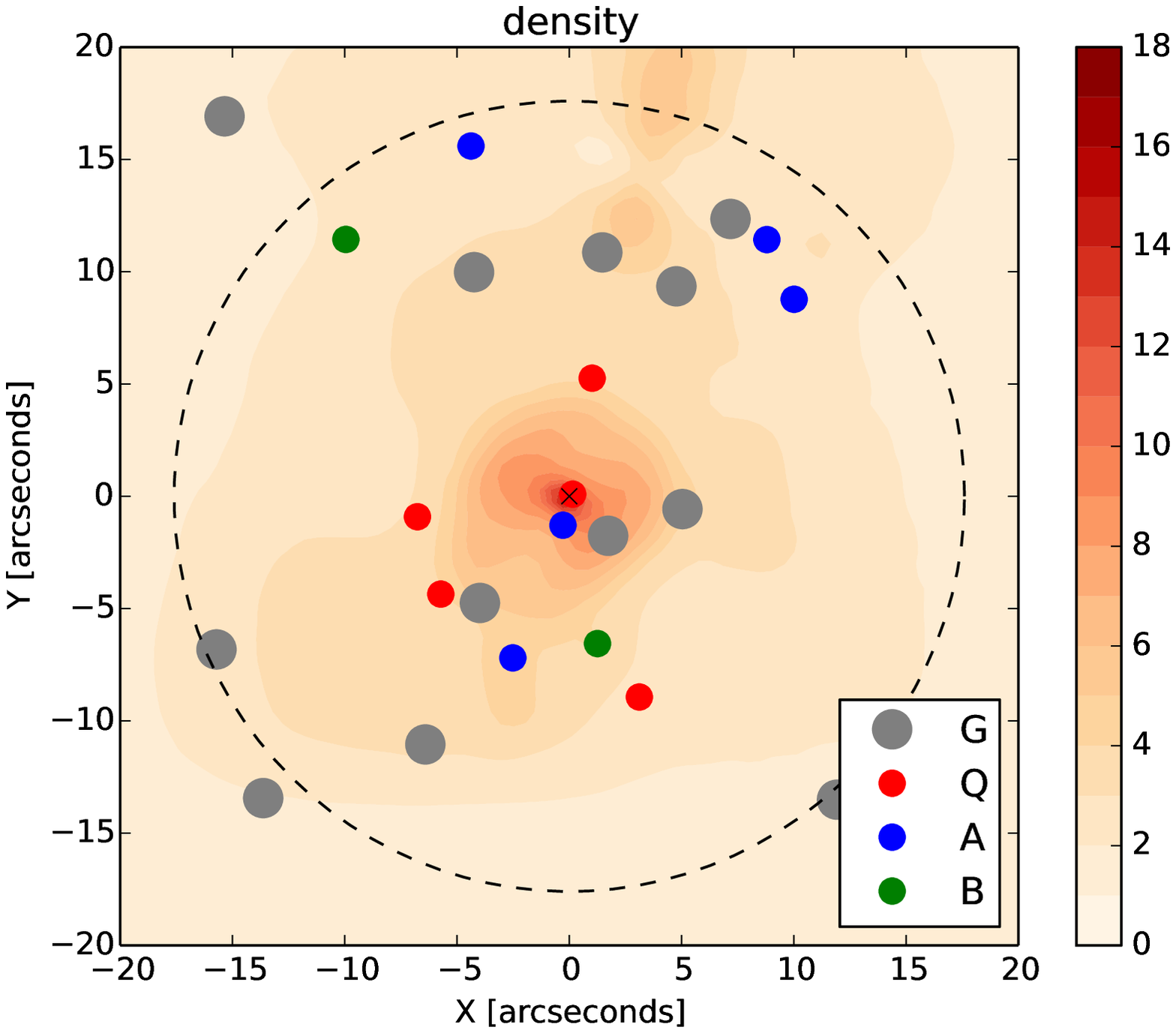}
\includegraphics[width=0.49\textwidth]{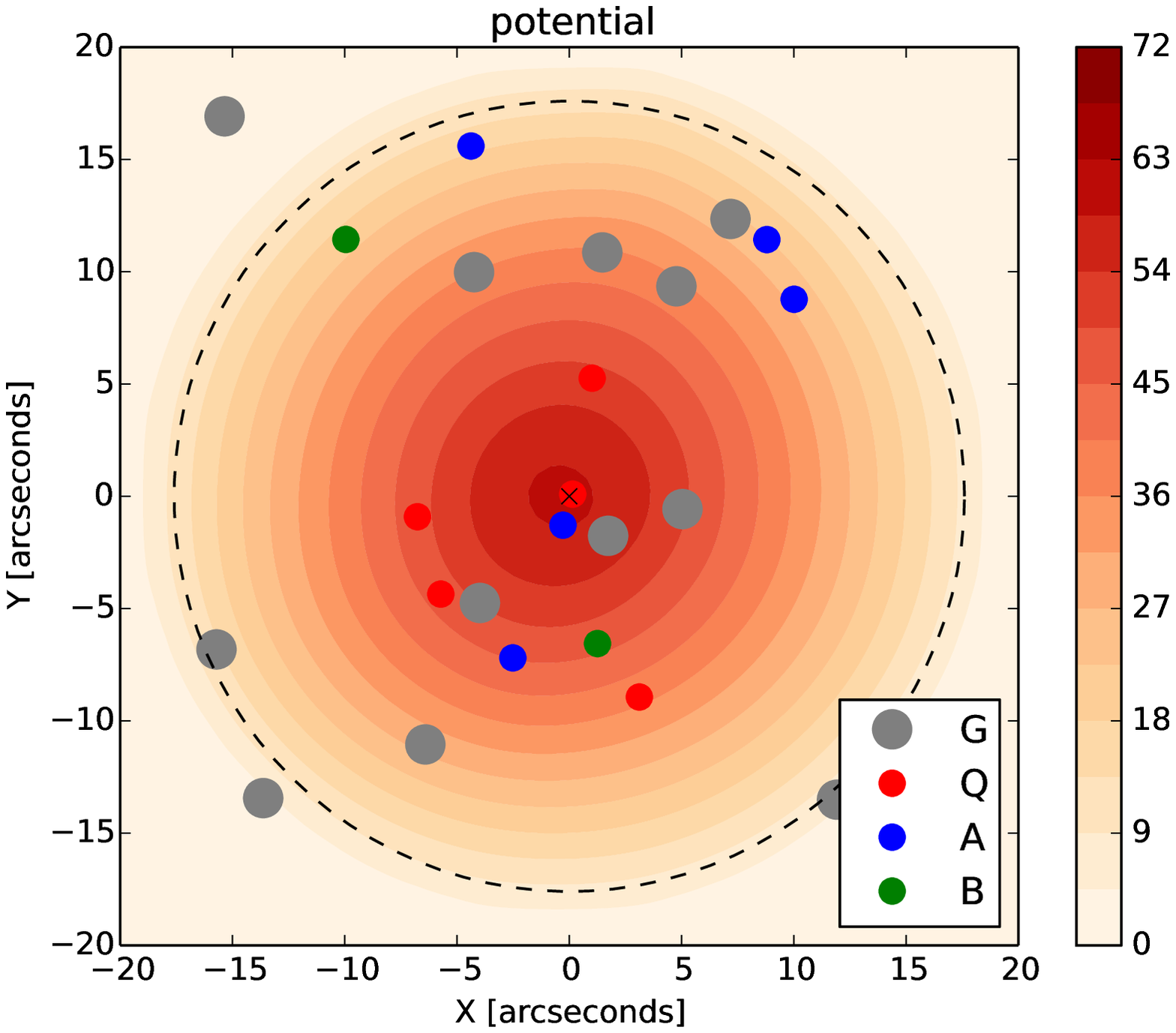}
\caption{The reference model (equation \ref{eq:refmodel}) using all
  the data constraints.  Density (left panel) means projected density
  in $\rm kg\,m^{-2}$ and potential (right panel) is $t_{\rm grav}$ in
  years.  Lensed images and cluster galaxies are marked.}
\label{fig:ref}
\end{figure*}

\section{Time delays in lensing}

Let us briefly recall the physics of time delays.  This is
conveniently done using Fermat's principle as applied to gravitational
lensing \citep{1986ApJ...310..568B}.  Consider a sky-projected density
$\Sigma(\vec x)$ at redshift $z_L$.  Here $\vec x$ represents the physical
(not comoving) coordinates perpendicular to the line of sight.  The
gravitational time delay due to this mass will be
\begin{equation}
\nabla^2 t_{\rm grav} = -(1+z_L)\frac{8\pi G}{c^3} \, \Sigma(\vec x\,) \,.
\end{equation}
Now consider a source at $\vec x_s$.  A light ray coming from this
source, being deflected at the lens so that it heads to the observer,
will have an additional geometrical time delay of
\begin{equation}
t_{\rm geom} = \frac{(1+z_L)}{2c} \frac{d_S}{d_L d_{LS}}
               \left| \vec x - \vec x_s\right|^2
\end{equation}
where the $d_L$ is the angular-diameter distance to the lens, and $d_S$
and $d_{LS}$ are angular diameter distances from observer to source
and from lens to source respectively.  In a flat cosmology
\begin{equation}
d_{z_1,z_2} = \frac{c/H_0}{1+z_2}\int_{z_1}^{z_2}
              \frac{dz}{\sqrt{\Omega_m(1+z)^3 + \Omega_\Lambda}}
\end{equation}
gives the various angular-diameter distances.

The total time delay is then
\begin{equation}
t(\vec x, \vec x_s) = t_{\rm geom}(\vec x,\vec x_s) + t_{\rm grav}(\vec x) \,.
\end{equation}
Images will form where $t(\vec x, \vec x_s)$ has a minimum, saddle
point, or maximum.  The tensor magnification is the inverse matrix of
second derivatives of $t(\vec x)$.  Note that the dependence on $\vec
x_s$ has been differentiated out.  Flexion consists of the derivatives
of the tensor magnification, hence third derivatives of the time.

Lens modelling consists of reconstructing $\Sigma(\vec x)$ and $\vec
x_s$.  For a quasar source, $\vec x_s$ is a single point, for an extended
source a superposition of source points must be considered.
The earliest detailed lens models \citep{1981ApJ...244..736Y} already
noted the non-uniqueness of lens models. \cite{1985ApJ...289L...1F}
quantified the most important of these, now known as the steepness
degeneracy or the mass-sheet degeneracy: steeper mass profiles give
longer time delays, while leaving image positions and shapes the
same. As modellers explored models further, it turned out that the
shape of the mass distribution also affects time delays
\citep{1997MNRAS.292..148S,2003ApJ...582....2Z,2006ApJ...653..936S}.
Degeneracies have also been studied theoretically \citep{2013A&A...559A..37S}.
Having sources at multiple different redshifts (high redshift
contrast) tends to suppress degeneracies
\citep{1998AJ....116.1541A,2009ApJ...690..154S} but does not eliminate
them completely \citep{2008MNRAS.386..307L,2012MNRAS.425.1772L}.

It is useful to break down lensing time delays into three factors:
lens substructure, lens size and cosmology.  The time delay between
the innermost and outermost images can be written as
\begin{equation}
   \Delta t_{\rm in,out} = f_{\rm lens} \frac{A_{\rm lens}}{A_{\rm
       sky}} \, \frac{d_Ld_S}{d_{LS}} \quad
A_{\rm lens} = \frac\pi4 (\theta_{\rm in}+\theta_{\rm out})^2 .
\end{equation}
Cosmology enters through the distance factors, while $A_{\rm lens}$ is
the size of the lens on the sky, and is fixed by the astrometry.  With
these factors fixed, $f_{\rm lens}$ is the remaining dependence on
substructure.  Typical values are 2--6 for systems with 2+1 images,
and 0.5--2 for systems with 4+1 images \citep{2006ApJ...650L..17S}.
That is to say, substructure is very important for time delays.  This
dependence is undesirable when estimating cosmological parameters, but
it is welcome for inferring substructure.

\section{Isolating the time-delay modes}\label{sec:theory}

In this section we produce a form of principal components analysis
(PCA) to isolate the information that time delays provide on the mass
distribution, assuming the cosmological parameters are known. A
somewhat related technique, for lensing clusters with multiple source
redshifts but not necessarily including time delays, is developed by
\cite{2014MNRAS.437.2461L}.

We reconstruct the mass distribution in two ways: first, including the
measured time delays (say TD models), and second, with no time-delay
information (say NTD models).  For each of TD and NTD, we
reconstructed an ensemble of mass maps (30 in number).  Each mass map
is on a grid (size $74\times 74$).

We now wish to find the variation present in the NTD ensemble but not
in the TD ensemble.  This will provide information on substructure
possibilities left by image data (only) but ruled out by time delays.
Let $X_n^i$ denote the projected density of the $i$th TD model ($i$
going from 1 to 30 in this paper) at the $n$th grid point (of
$74^2=5476$ grid points).  Similarly use $Y_n^i$ for the NTD mass
maps.  Next, we choose a reference $Z_n$,
\begin{equation} \label{eq:refmodel}
  Z_n = \langle X_n^i \rangle
\end{equation}
which is the ensemble average of the TD maps.  Then
\begin{equation}
  \Delta X_n^i = X_n^i - Z_n
\end{equation}
is the ensemble variation about the reference.  Then we introduce
a moment matrix
\begin{equation}
  M_{mn}(X) = \left\langle \Delta X_m^i \, \Delta X_n^i \right\rangle
\end{equation}
where the average is again over the TD maps.  $M_{mn}(X)$ is just the
covariance between pairs of grid points.  The eigenvalues and
eigenvectors of $M_{mn}(X)$ describe how sets of grid points tend to
vary together.  Let us denote them by $\lambda_k(X)$, and $V_n^k(X)$
respectively; the superscript $k$ denotes the $k$th eigenvalue and
eigenvector ($k=1$ having the largest eigenvalue).  In practice, the
first few eigenvalues dominate.  The vector
\begin{equation}\label{eq:TD}
   Z_n \pm \sqrt{|\lambda_k(X)|} \, V_n^k(X),
\end{equation}
displayed as a density map, shows the principal mode of variation of
the mass model.  (There is no sum over $k$.)  These variation modes
are, of course, orthogonal.

We then proceed to the NTD maps. Let these be $Y^i_n$ and let
\begin{equation}
  \Delta Y_n^i = Y_n^i - Z_n
\end{equation}
be the variations with respect to the reference model.  We now
subtract off the variation modes of TD mass maps
\begin{equation}
   \Delta \bar Y_n^i = \Delta Y_n^i - \sum_{k}
   \left( {\textstyle\sum_m} \Delta Y^i_m V_m^k(X) \right) V_n^k(X)
\end{equation}
leaving variations that are orthogonal to the TD
variations.  Using these, we build another moment matrix
\begin{equation}
  M_{mn}(Y) = \left\langle \Delta \bar Y_m^i \, \Delta \bar Y_n^i \right\rangle.
\end{equation}
The eigenvalues and eigenvectors of $M_{mn}(Y)$ contain the modes
present in NTD maps models but absent from the TD maps.  These modes
can be conveniently displayed as
\begin{equation}\label{eq:NTD}
   Z_n \pm \sqrt{|\lambda_k(Y)|} \, V_n^k(Y).
\end{equation}
We call these ``variations ruled out by the time delays'' and these
are the main results of this paper.  One sign of the $\pm$ must be
chosen for definiteness, but it does not matter which one.

\begin{figure}
\centering
\includegraphics[width=0.49\textwidth]{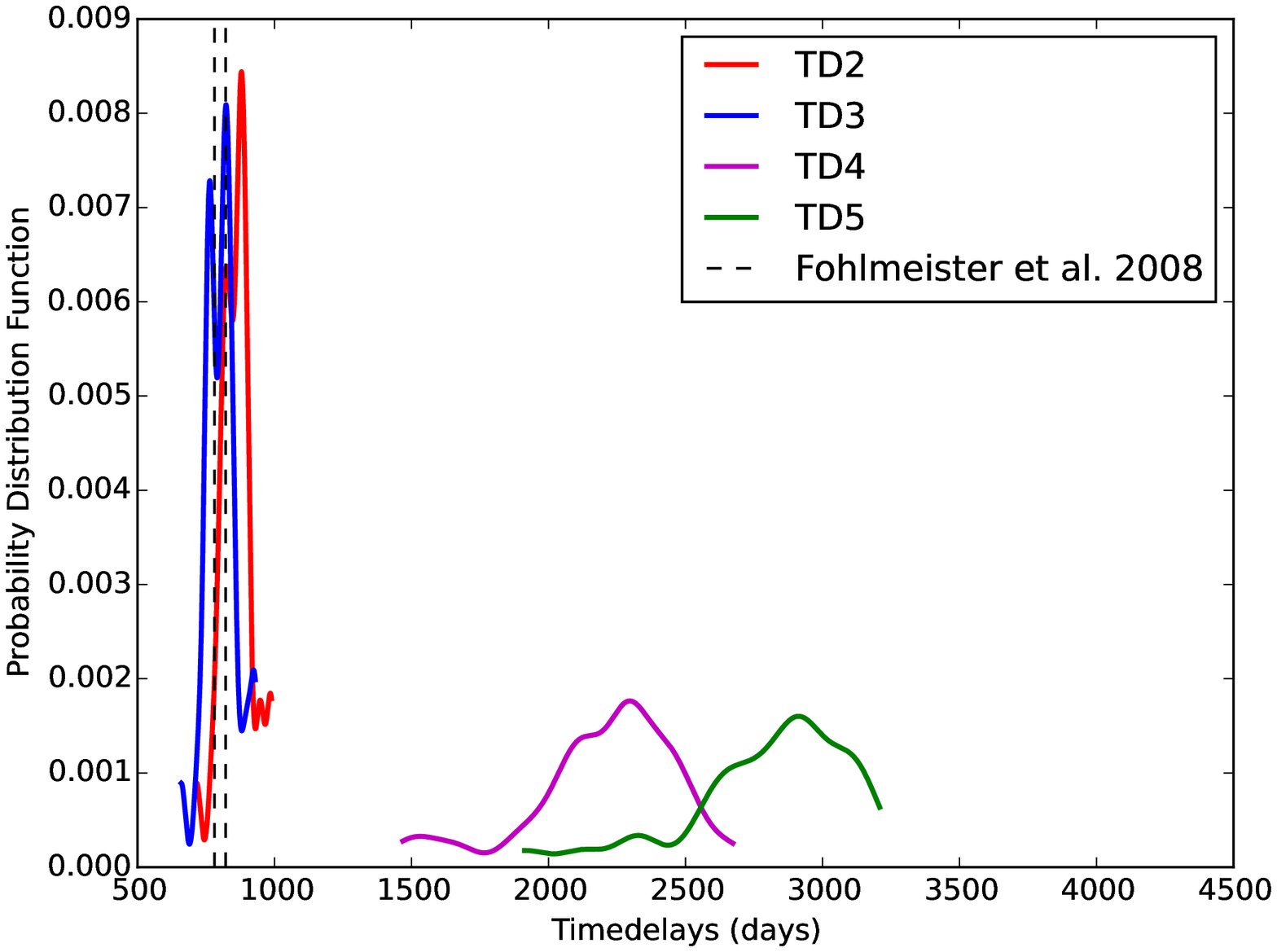}\\
\includegraphics[width=0.49\textwidth]{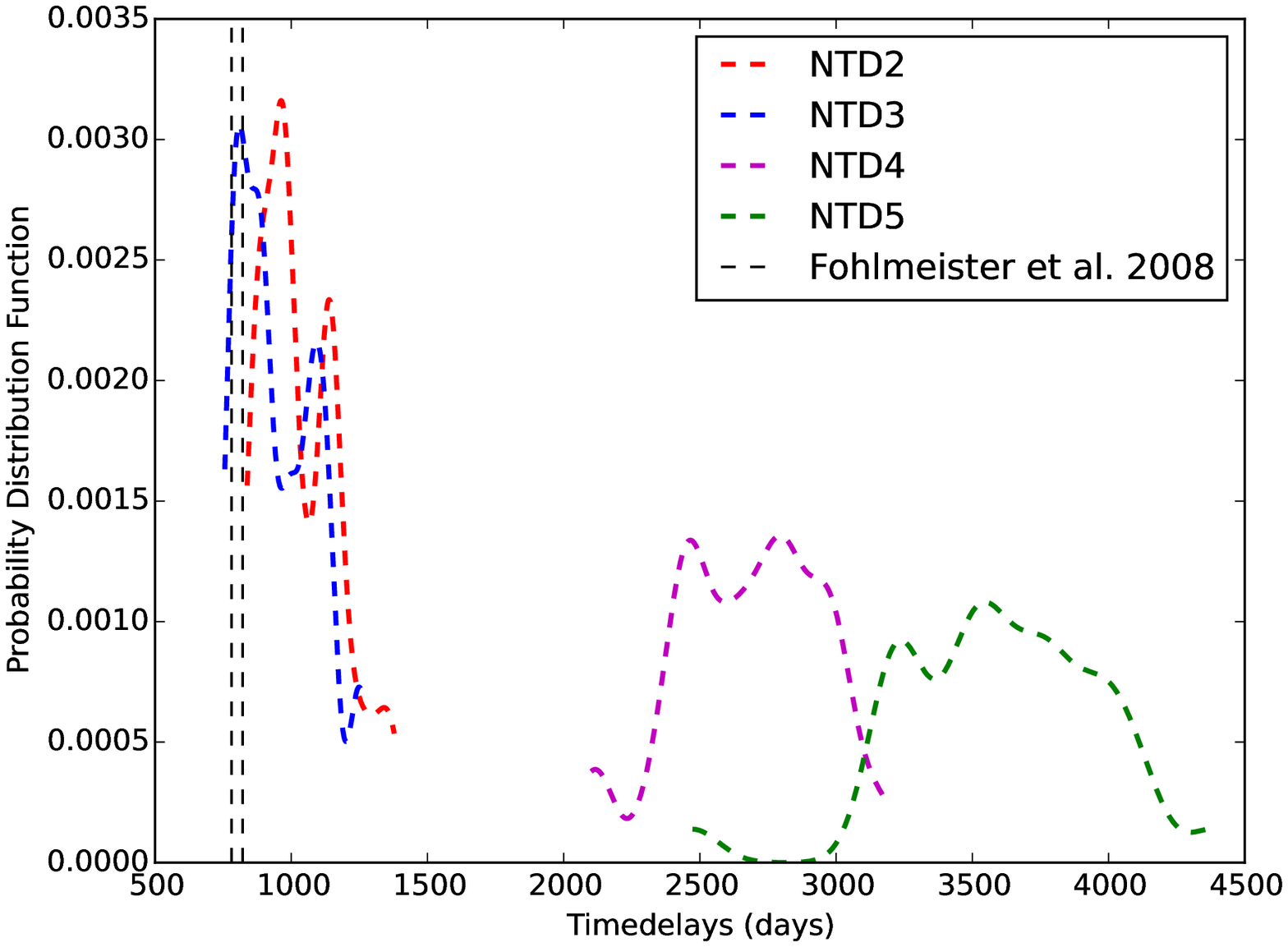}
\caption{Time delay estimates using TD models (upper) and NTD models
  (lower).  TD2, TD3, TD4 and TD5 (or NTD2, NTD3, NTD4, NTD5) are 
  the time delays for image Q2, Q3, Q4 and Q5
  respectively with respect to Q1.  Vertical dashed lines show the
  input time delays from \cite{2008ApJ...676..761F}.}
\label{fig:timedelays}
\end{figure}

\section{Application to SDSS J1004+4112}

The mass distribution $\Sigma(\vec x)$ can be reconstructed in
different ways: \cite{2010PASJ...62.1017O} uses a parametrized form,
\cite{2007ApJ...663...29S} built it out of mass tiles or pixels, while
\cite{2009MNRAS.397..341L} used an adaptive superposition of Plummer
components.  The latter, and in particular the GRALE code, is used
in the present work.

\subsection{Mass reconstruction using GRALE}\label{sec:grale}

As mentioned above, GRALE
\citep{2006MNRAS.367.1209L,2007MNRAS.380.1729L,2009MNRAS.397..341L}
makes free-form mass models for the lensing cluster, as a
superposition of many Plummer lenses.  Except for the redshift and 
general location, no information from the lens itself is used.
The inversion input consists of (1)~lensed-image
positions and source redshifts, (2)~regions where additional images
have not been identified but could be present, and (3)~time delays if
any.  Each of these is used to define a fitness measure of a mass
distribution.

\begin{itemize}
  \item For a given mass map, the input multiple images are ray-traced
    back to the source plane.  The `overlap fitness' of the mass map
    expresses how well these back-projected images overlap.  It is
    important to consider fractional overlap rather than simple
    source-plane distances to avoid favouring extreme magnification
    (tiny sources).
  \item The back-projection may give rise to extra images.  If they
    are not in regions specified by the modeller as allowed, they are
    spurious. These penalize the mass map through a `null fitness'
    measure.  Through the null fitness, GRALE uses the non-occurrence
    of images at random locations as useful data.
  \item The `time-delay' fitness measures how well the time delay in a
    mass distribution agrees with the observations.
\end{itemize}

GRALE uses a multi-objective genetic algorithm to find free-form mass
maps which provide optimal fits to the data according to the above
criteria.  If more Plummer-lens components are allowed, the fitness
will tend to be better.  Accordingly, we used a heuristic
Occam's-razor criterion to find a compromise between better fitness
and more components \citep{2014MNRAS.439.2651M}.  This effect sets the
resolution adaptively.  No additional priors or regularization are
used.

\subsection{Mass models}

Our mass maps for SDSS J1004+4112 were of two kinds, as follows.
\begin{itemize}
\item No time delay (or NTD) models.  We used the three image systems
  (Q, A and B; total 12 images), but gave no time-delay information in
  this case. GRALE finds an optimal solution, restricting extra images
  using the technique of null spaces, for this dataset. We repeated
  the same procedure 30 times to generate a statistical ensemble of 30
  models.
\item Time delay (or TD) models.  For these we used the same data set,
  plus the two measured time delays: 40 days in Q2--Q3 and 821 days in
  Q1--Q3 (see Figure \ref{fig:raw}).  Following the same procedure, we
  made an ensemble of 30 mass maps. Figure \ref{fig:ref} shows the
  ensemble average of the TD models, which is used as the reference
  $Z_n$ from Section \ref{sec:theory}. The left panel of Figure
  \ref{fig:ref} shows the projected density and the right panel shows
  the potential in the colour code. This mass distribution shows slight
  differences from \cite{2009MNRAS.397..341L}, because the present
  work uses an adaptive resolution scheme, but these differences do
  not appear to be significant.
\end{itemize}

The method described in Section \ref{sec:theory} is then applied to
the TD and NTD models.  Tests in \cite{2014MNRAS.439.2651M} indicate
that GRALE ensembles of this size underestimate the actual
uncertainties by a factor of two, but do explore the different
uncertainties.  Thus, the eigenvalues reported below are certainly
underestimates, but the eigenvectors should be a good representation
of the variation.

\begin{figure}
\centering
\includegraphics[width=0.49\textwidth]{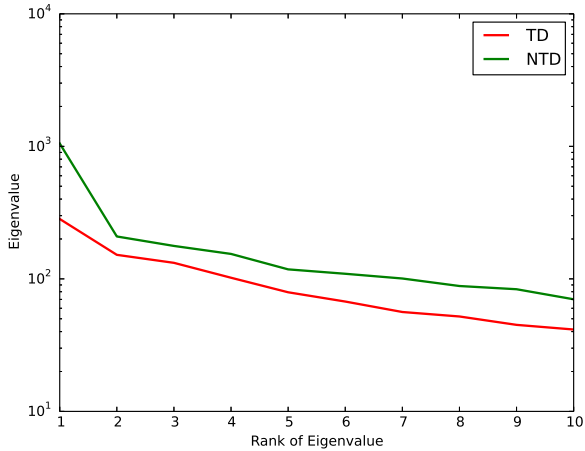}
\caption{Distribution of eigenvalues $\lambda_k(X)$ and
  $\lambda_k(Y)$, labelled as TD and NTD respectively, ranked by
  absolute value.}
\label{fig:eigvals}
\end{figure}

\begin{figure*}
\centering
\includegraphics[width=0.99\textwidth]{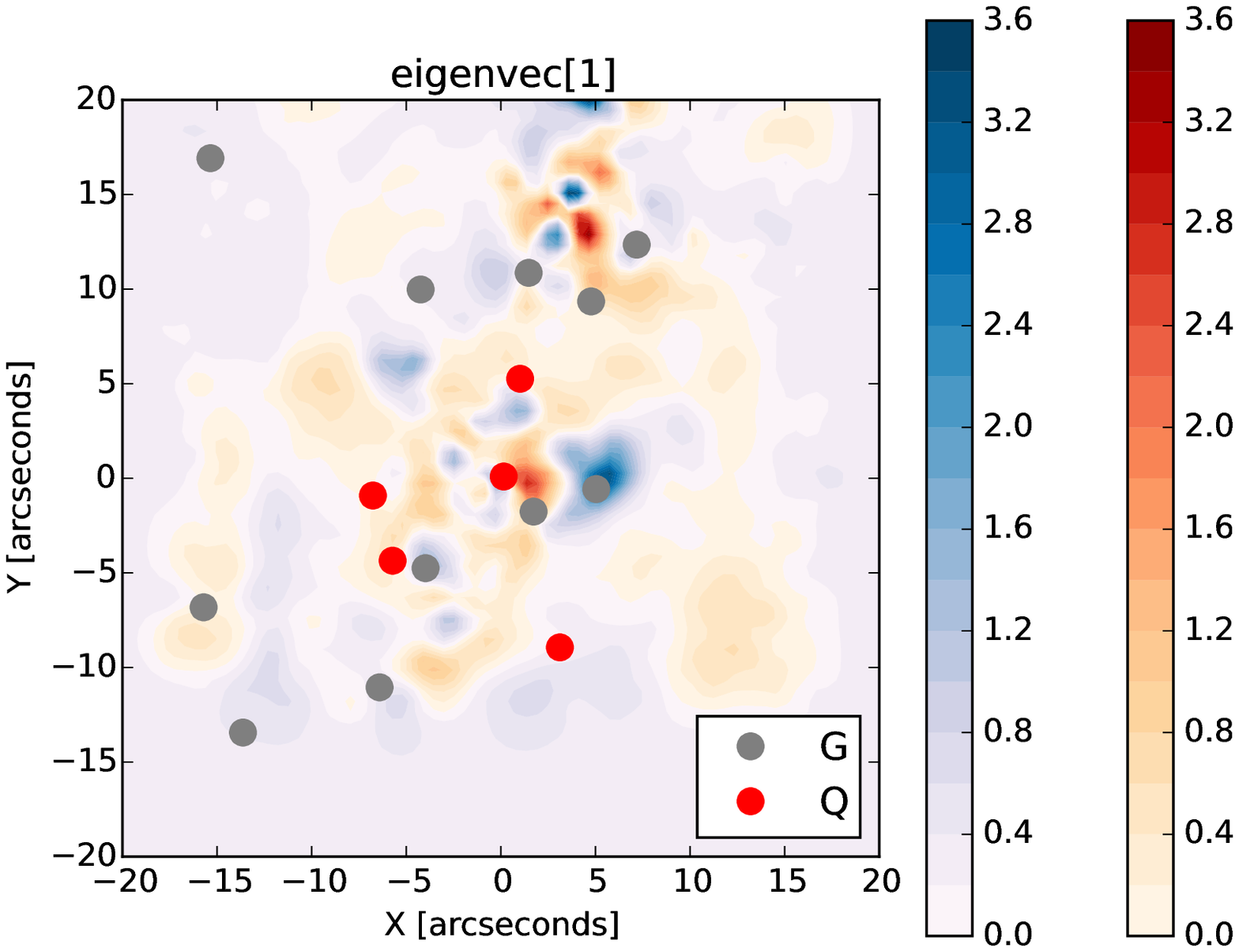}
\includegraphics[width=0.49\textwidth]{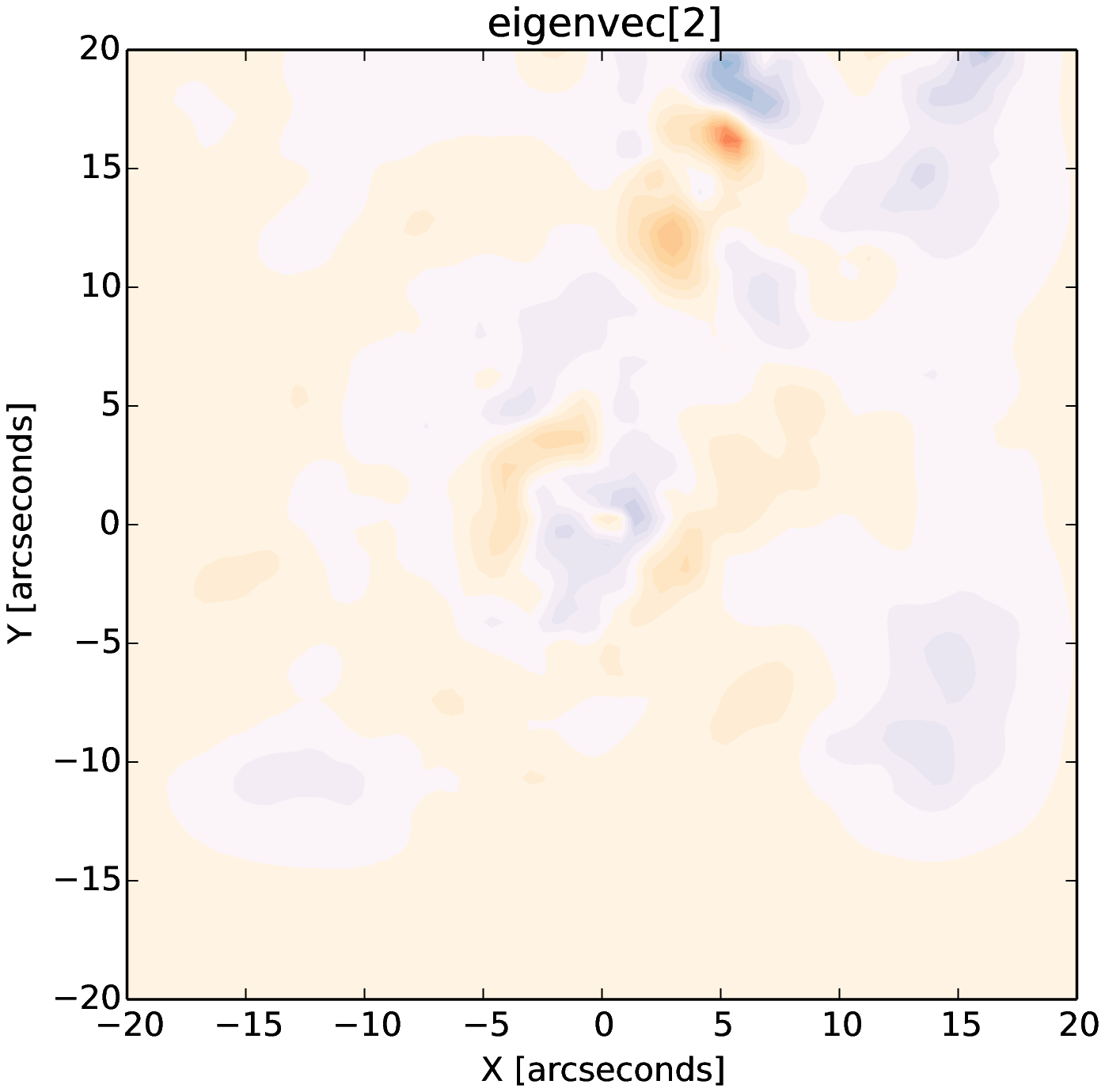}
\includegraphics[width=0.49\textwidth]{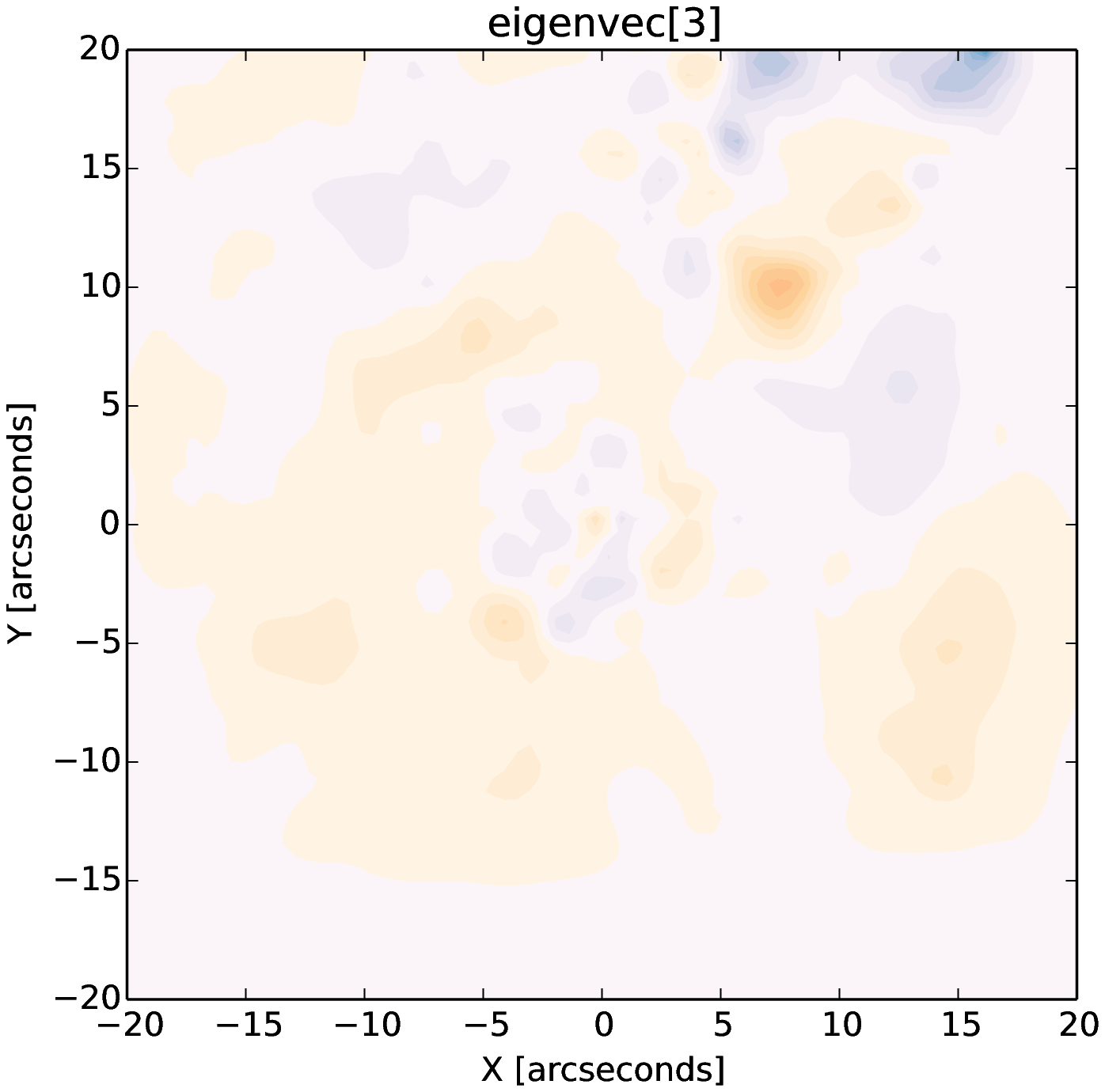}

\caption{The main/largest variation in surface density
  ruled out by time delays (upper
  panel). The red and the blue color bars represents positive and
  negative values (in $\rm kg\,m^{-2}$) respectively.  The two lower
  panels show the second and third largest variation ruled out by time
  delays.}
\label{fig:mainmode}
\end{figure*}

\begin{figure*}
\centering
\includegraphics[width=0.99\textwidth]{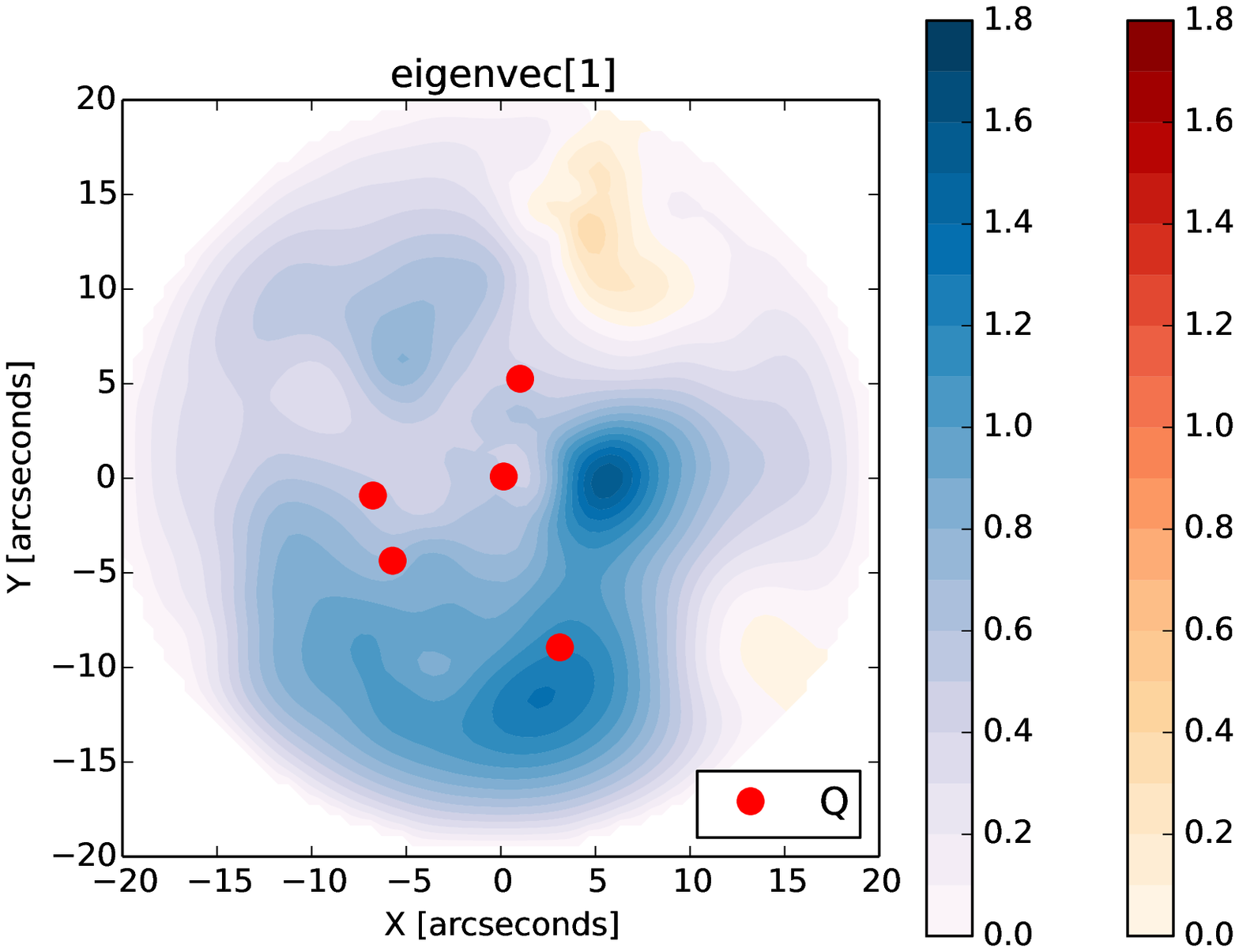}
\caption{The main/largest variation ruled out by time delays. The red
and the blue colour bars represents positive and negative values (in years) respectively. }
\label{fig:mainmode-potential}
\end{figure*}

\subsection{Results and interpretation}

Figure \ref{fig:timedelays} shows the distribution of time delays in
the TD and NTD ensembles.  As expected in the TD models, the first two
time delays, which were also used as data inputs, show little
variation (though larger than the observational uncertainties) while
the other two show large variations.  In the NTD models, all the time
delays show large variations.  Most of the TD models also favour lower
time delays for Q4 and Q5, as compared to the NTD models estimate.  Thus,
existing time-delay measurements constrain future measurements to some
degree, but not very tightly, indicating that future measurements will
bring substantial new information.

Figure \ref{fig:eigvals} shows the eigenvalues of the NTD and TD
modes.  The largest NTD mode evidently dominates, being a factor of
five larger than the second largest mode.  When time delays are
included, it is also the case that one TD mode is much larger than the
rest.

\begin{figure*}
\centering
\includegraphics[width=0.49\textwidth]{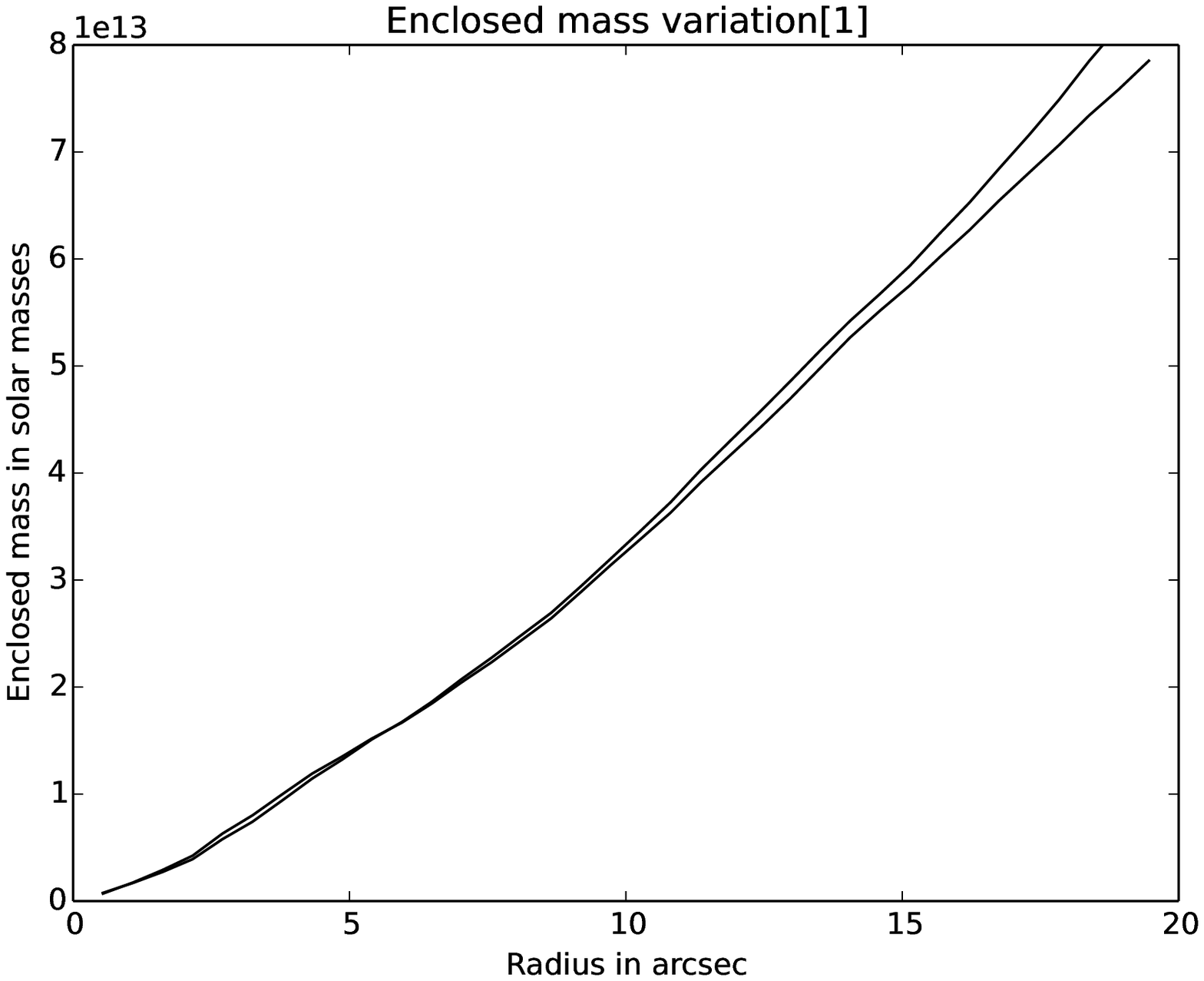}
\includegraphics[width=0.49\textwidth]{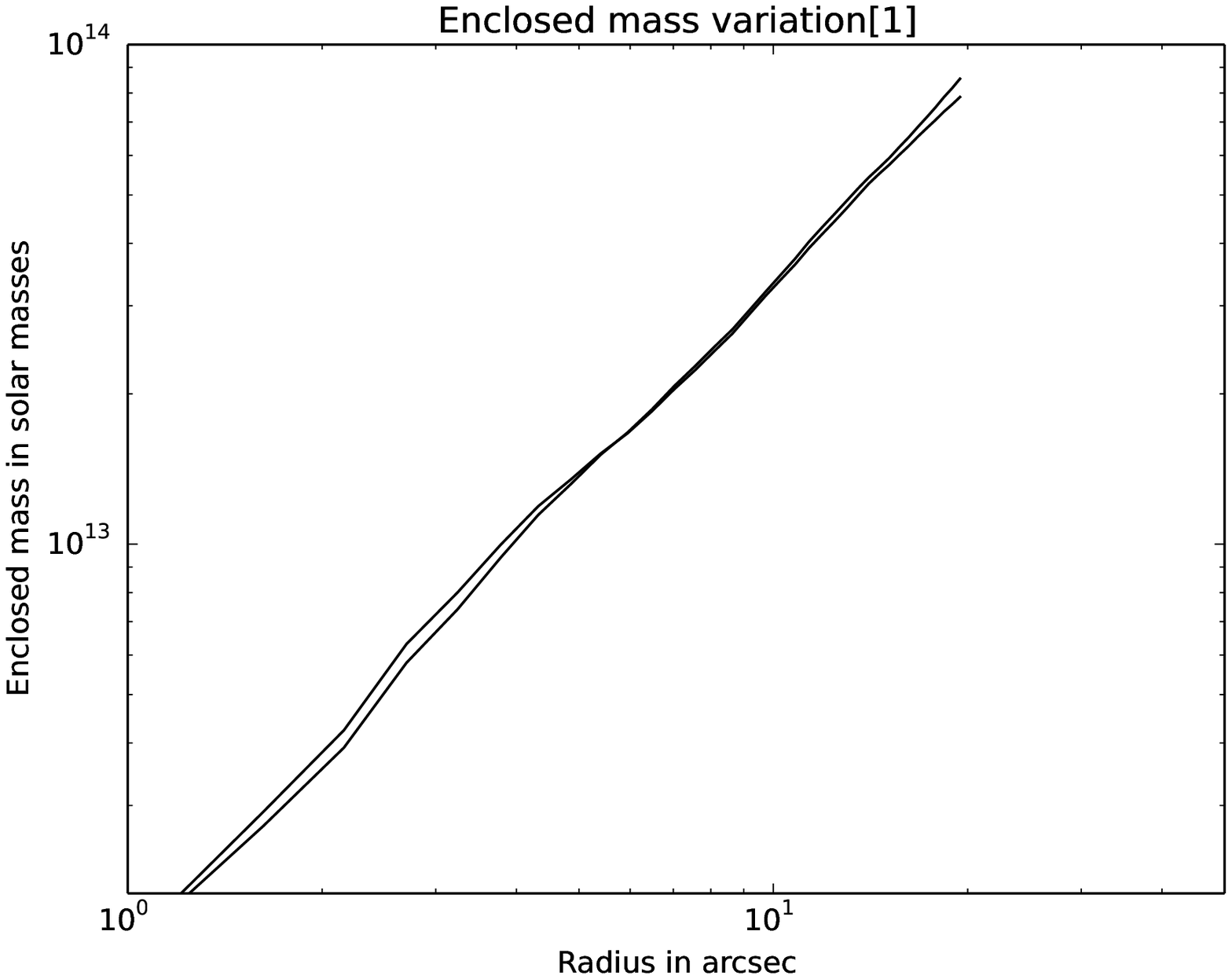}
\caption{Variation of enclosed mass of the reference model around the 
largest NTD mode (shown in Figure \ref{fig:mainmode}) as a function of 
projected radius.  Linear scales on the left, log scales on the right.}
\label{fig:mainmode-menc}
\end{figure*}

Figure \ref{fig:mainmode} presents the main result of this paper.  It
illustrates the largest NTD variation mode that is absent in the TD
ensemble (cf.\ equation \ref{eq:NTD}), in other words, the largest
variations ruled out by the time-delay measurements.  The upper panel
of the figure shows the eigenvector corresponding to the largest
eigenvalue.  The two central blobs in the upper panel of Figure
\ref{fig:mainmode} (blue and red, hence anti-correlated) allow the
mass to shift the local mass peak towards or away from the cluster
centre, and hence change the steepness of the mass profile. Without
the TD constraints, the central peak of the mass distribution can be
 more lopsided, whereas the TD constraints force the central peak to
be rounder.  The rather large uncertainties (20\%--40\% in $\Sigma$)
in the central region of the NTD models is reduced in the TD
models. This local uncertainty of 20\%--40\% is, however, less than
1\% of the total mass in the strong-lensing region.  That is to say,
time delays are constraining substructures that are only a percent of
the total.  The red and blue blobs also are in the vicinity of
the two galaxies in the very central region.  That may be coincidental,
but it is worth remarking that the mass in the blobs is of the order of a
galaxy-halo mass, such as may be tidally stripped from a galaxy near
the centre of a cluster.  This suggests that time delays may give very
sophisticated information about the variation of the mass near the
environment of the galaxies in the central region, or in other words,
about the sub-structures in a dense environment. Therefore, the third
time-delay, which is expected for the image Q4, could be very useful
in extracting the substructure information of the cluster near its
centre.  Note that this interesting region is not accessible through
the more well-known techniques of weak lensing and flexion.

The lower two panels of Figure \ref{fig:mainmode} shows the second and
third largest variation modes. These variations modes are weaker,
also evident from Figure \ref{fig:eigvals} as all other
eigenvalues after the first one are sub-dominant.  The number of
non-zero eigenvalues for NTD should equal the number of time-delay
measurements (two in this case), but in practice further modes are
present, and only gradually die away.  These are basically noise
modes, which exist because we have only 30 models for our
principal-components analysis.  (Numerical noise due to round-off
error in the matrix operations is negligible.)

Upon measuring the short Q2-Q3 time delay \cite{2007ApJ...662...62F}
already noted that substructure would be necessary to account for
their observation.  We have not considered separately the case where
only this one time delay is known.  We see in
Figure~\ref{fig:eigvals}, however, that only one of the NTD
eigenvalues is larger than the TD eigenvalues, indicating that only
one of the measured time delays (surely the longer Q1-Q3 value) gives
a substructure constraint.  This does not mean that the Q2-Q3
measurement can be explained without substructure; only that the
measurement on its own is not constraining what that substructure
could be.

Figure \ref{fig:mainmode-potential} shows the variation in lens
potential $t_{\rm grav}$ corresponding to the density variation from
the main panel in Figure \ref{fig:mainmode}.  Here we can see the
effect of measuring the delays between Q1, Q2, and Q3.  The NTD models
have $t_{\rm grav}$ between Q1 and Q2 varying by about 0.5 years, that
is, varying by 25\% of the measured value.  Between Q2 and Q3, the
time delay varies by about 0.3 years in the NTD models, which is more
than twice the measured delay between this pair.  Between Q3 and Q4
there is little variation.  This does not mean that the NTD models
have little variation in the Q3--Q4 delay; it just means that similar
variation is present in the TD models, and hence not ruled out by the
Q1--Q2--Q3 measurements.  All this is what one would have expected
from Figure \ref{fig:timedelays}.  Surprising, however, is that the
largest variation ruled out by the time delays is the blob $5''$ west
of the cluster centre, not near any of the images.

Figure \ref{fig:mainmode-menc} shows the enclosed mass and its
variation.  The enclosed mass is comparable with Figure~3 of
\cite{2010PASJ...62.1017O}, as expected.  As also evident from figure
\ref{fig:mainmode}, the NTD mode is not like a global change of steepness --- the
steepness appears to have been already constrained by the image data,
because of the redshift contrast between the Q, A and B systems.  So
the time delays are giving us information not on the steepness but
about the shape of the profile.  The NTD models allow for strongly E/W
ellipticity in the central region, changing to a more N/S elongation
further out, but the time delays force the model to be rounder and
less lopsided.

\section{Discussion}

This paper expresses the information that comes from lensing time
delays in a way that is orthogonal to other lensing observables.  We
used the cluster SDSS J1004+4112 because of the richness of
strong-lensing information.  SDSS J1029+2623
\citep{2013ApJ...764..186F,2013MNRAS.429..482O} would also be
interesting for a similar study, as would MACS J1149+2223
\citep{2014arXiv1411.6009K,2014arXiv1411.6443O} if time delays for the
recently-discovered supernova can be measured.  The main results are
shown in figures \ref{fig:mainmode} and \ref{fig:mainmode-potential}.
The interpretation is very preliminary, because a new time delay is
expected soon, but nonetheless shows two interesting features.

First, it is remarkable how a small mass redistribution can produce a
large difference in time delays.  As Figure \ref{fig:mainmode} shows,
the main mass-redistribution ruled out by the two published time
delays is a small blob that is only $\sim1\%$ of the total mass in the
strong-lensing region, and yet this redistribution can change a time
delay by 50\%.  It appears that time delays are providing substructure
constraints at the 1\% level.  

Second, the mass redistribution is not mainly near the images, but
mainly in another region of the cluster.  Overall, the time delays
reduce the allowed lopsidedness of the cluster, but it is intriguing
that the main redistribution appears to shift mass from the
neighbourhood of one cluster galaxy to the vicinity of another cluster
galaxy.

The expected third time-delay measurement in this cluster will be very
interesting. As seen in Figure \ref{fig:raw}, the two time delays we used
in our analysis are between Q1--Q2 and Q1--Q3. All of these images are
towards the west and south-east of the cluster.  The new measurement
would involve Q4, which lies to the north.

So, we conclude that strong lensing with time delays provides important
constraints on the distribution of matter near the centre of the
lensing cluster, regions not accessible to weak lensing or flexion.
Additionally, time delays may provide new information on the 
distribution of matter in the
densest regions of clusters and indirectly on the role of AGN feedback,
adiabatic contraction, and other dynamical processes.

\bibliographystyle{mn2e}

\def\apj{ApJ}
\def\apjl{ApJL}
\def\aj{AJ}
\def\mnras{MNRAS}
\def\aap{A\&A}
\def\nat{Nature}
\def\pasj{PASJ}

\bibliography{ms.bib}

\end{document}